# Frequency-independent terahertz anomalous Hall effect in DyCo$_5$, Co$_{32}$Fe$_{68}$ and Gd$_{27}$Fe$_{73}$ thin films from DC to 40 THz

*Tom S. Seifert\*, Ulrike Martens, Florin Radu, Mirkow Ribow, Marco Berritta, Lukáš Nádvorník, Ronald Starke, Tomas Jungwirth, Martin Wolf, Ilie Radu, Markus Münzenberg, Peter M. Oppeneer, Georg Woltersdorf, Tobias Kampfrath*

Dr. Tom S. Seifert, Prof. Dr. Tobias Kampfrath
Department of Physics, Freie Universität Berlin, 14195 Berlin, Germany
Department of Physical Chemistry, Fritz-Haber-Institute of the Max-Planck-Society, 14195 Berlin, Germany
E-mail: tom.seifert@fu-berlin.de

Dr. Ulrike Martens, Prof. Dr. Markus Münzenberg
Institute of Physics, University of Greifswald, 17489 Greifswald, Germany

Dr. Florin Radu
Helmholtz-Zentrum Berlin für Materialien und Energie, Albert-Einstein-Str. 15, 12489 Berlin, Germany

Dr. Mirkow Ribow, Prof. Dr. Georg Woltersdorf
Institute of Physics, Martin-Luther Universität Halle-Wittenberg, 06120 Halle (Saale), Germany

Dr. Marco Berritta, Prof. Dr. Peter M. Oppeneer
Department of Physics and Astronomy, Uppsala University, P.O. Box 516, SE-75120 Uppsala, Sweden

Dr. Lukas Nadvornik
Faculty of Mathematics and Physics, Charles University, Ke Kalovu 2027/3, Prague 12116, Czech Republic

Prof. Dr. Tomas Jungwirth
Institute of Physics, Czech Academy of Sciences, Cukrovarnicka 10, 162 00, Praha 6, Czech Republic
School of Physics and Astronomy, University of Nottingham, NG7 2RD, Nottingham, United Kingdom

Dr. Ronald Starke
TU Bergakademie Freiberg, 09599 Freiberg, Germany

Prof. Dr. Martin Wolf
Department of Physical Chemistry, Fritz-Haber-Institute of the Max-Planck-Society, 14195 Berlin, Germany

Dr. Ilie Radu
Max-Born Institute for Nonlinear Optics and Short Pulse Spectroscopy, Max-Born-Str. 2A, 12489 Berlin, Germany
Department of Physics, Freie Universität Berlin, 14195 Berlin, Germany

Keywords: terahertz time-domain spectroscopy, terahertz spintronics, terahertz ellipsometry, spintronic terahertz emitter

The anomalous Hall effect (AHE) is a fundamental spintronic charge-to-charge-current conversion phenomenon and closely related to spin-to-charge-current conversion by the spin Hall effect. Future high-speed spintronic devices will crucially rely on such conversion effects at terahertz (THz) frequencies. Here, we reveal that the AHE remains operative from DC up to 40 THz with a flat frequency response in thin films of three technologically relevant magnetic materials: $DyCo_5$, $Co_{32}Fe_{68}$ and $Gd_{27}Fe_{73}$. We measure the frequency-dependent conductivity-tensor elements $\sigma_{xx}$ and $\sigma_{yx}$ and find good agreement with DC measurements. Our experimental findings are fully consistent with *ab-initio* calculations of $\sigma_{yx}$ for CoFe and highlight the role of the large Drude scattering rate (~100 THz) of metal thin films, which smears out any sharp spectral features of the THz AHE. Finally, we find that the intrinsic contribution to the THz AHE dominates over the extrinsic mechanisms for the $Co_{32}Fe_{68}$ sample. The results imply that the AHE and related effects such as the spin Hall effect are highly promising ingredients of future THz spintronic devices reliably operating from DC to 40 THz and beyond.

# 1. Introduction

Incorporating the electron spin into electronic devices is the central idea of spintronics.[1] This growing research field ultimately aims at generating, controlling and detecting spin currents at terahertz (THz) rates.[2] To realize such high-speed spin operations, spin-orbit interaction (SOI), despite being weak, plays a key role because it couples the motion of an electron to its spin state.[3] From a classical viewpoint, SOI can be understood as a spin-dependent effective magnetic field that deflects copropagating spin-up and spin-down conduction electrons in opposite directions (see **Figure 1a**). Important consequences of SOI are the spin Hall effect (SHE)[4] and its magnetic counterpart, the anomalous Hall effect (AHE).[5,6] In a metal with SOI, the SHE converts a charge current into a transverse pure spin current. Similarly, the AHE in a ferromagnetic metal causes a transverse spin-polarized charge current proportional to the net magnetization.[7]

Such SOI-induced effects have found broad application in spintronic devices for spin-current generation and detection as well as for switching of magnetic order.[8,9] Up to now, however, most spintronics work has been limited to frequencies below 10 GHz,[10] significantly lagging behind other information carriers such as electrons in field-effect transistors featuring cut-off frequencies of $\sim$ 1 THz.[11] Therefore, the question arises how SOI-induced effects evolve at THz frequencies. Previous ultrafast works demonstrated that the inverse SHE is still operative up to 30 THz.[12,13,14,15,16,17,18,19,20,21] However, its actual strength, in particular in comparison to low frequencies down to DC, is an open question. Its answer is highly relevant for the transfer of spintronic functionalities to the THz range[10,22], which can provide access to collective spin dynamics at their natural frequencies, including exchange modes in ferrimagnets[23] and antiferromagnets.[24,25]

From a fundamental viewpoint, studying THz spin-to-charge conversion yields insights into the energetic structure of SOI because the photon energy (4 meV at 1 THz) is comparable to typical SOI energy scales in solids. Since pure spin currents are much more difficult to measure than charge currents, it is reasonable to start with studying the THz AHE. So far, however, no AHE data are available over the entire range from 0 to about 100 meV for magnetic metals relevant to THz spintronics. Notable exceptions are measurements below 6 THz on $SrRuO_3$ (Ref. 26), magnetic semiconductors[27,28] and metals.[29,30] For infrared frequencies above 25 THz, again $SrRuO_3$ (Refs. 31, 32) and related compounds were studied.[33]

In this work, we use broadband THz time-domain ellipsometry in combination with DC AHE measurements to extract the complex in-plane conductivity tensor of magnetic metals from 0 to 40 THz, thereby closing the gap between DC and optical frequencies (Figure 1b). A comparison to *ab-initio* calculations suggests that the large electron scattering rate has two important consequences: First, it makes the THz AHE largely frequency-independent. Second, it reinforces the intrinsic AHE contribution.

We investigate magnetic metals representative of a whole class of materials with large SOI that become increasingly important in ultrafast spintronics:[34,35,36] ferromagnetic CoFe and the ferrimagnets DyCo$_5$ and GdFe.[37,38,44] Their potential ultrafast applications require characterization and understanding of the spintronic phenomena at accordingly high, that is, THz frequencies. We envisage that our novel broadband THz time-domain ellipsometry does not only allow us to study the presented, technologically highly-relevant materials but will also enable studies of the THz spintronic response of emerging material classes in the future.

## 2. Experimental details

### 2.1 Conceptual idea

In a DC AHE measurement (Figure 1a), an electrical voltage drives a spin-polarized current through a magnetic conductor with out-of-plane magnetization. SOI deflects spin-up and spin-down conduction electrons in opposite directions perpendicular to the sample magnetization and the driving current. The resulting transverse spin-polarized anomalous Hall current is measured electrically, usually limited to gigahertz frequencies. [39]

To cover the THz frequency range, we use a quasi-optical and contactless scheme (Figure 1b). A linearly polarized THz electric-field pulse drives a spin-polarized in-plane current in the magnetic metal film. The SOI-induced perpendicular anomalous Hall current emits THz radiation into the far-field. Consequently, the transmitted THz pulse becomes elliptically polarized. Using broadband THz time-domain ellipsometry, we measure the driving and induced THz electric field from 1 to 40 THz.

We note that the AHE is determined by the same conductivity tensor as the Faraday effect. Thus, our THz AHE measurement can also be considered as the THz Faraday effect, which is a more commonly used term at optical frequencies. Unlike with optical frequencies, our scheme allows us to directly study the Drude-response of the spintronically relevant conduction electrons close to the Fermi energy. For photon energies in the mid-infrared region (above ~0.1 eV), however, the free-carrier-like dynamics is possibly increasingly superimposed by interband transitions.[40]

### 2.2 Materials

We study two crystalline (CoFe and DyCo$_5$) and one amorphous (GdFe) material. All samples have an out-of-plane magnetic anisotropy, perfectly suited to achieve large THz AHE signals.

The ferromagnetic Co$_{20}$Fe$_{60}$B$_{20}$ film with the layer stacking MgO(2 nm)|Co$_{20}$Fe$_{60}$B$_{20}$(1 nm)|Ta(8 nm)||Si$_3$N$_4$(150 nm) was prepared by magnetron sputtering and electron-beam evaporation (see Supporting Information S3). The Fe-rich composition of CoFeB was chosen to ensure, on one hand, an out-of-plane magnetic anisotropy even at a thickness as large as 1 nm.[41] On the other hand, the Co content enhances the magnetic moment and, thus, the AHE signal. As prepared, Co$_{20}$Fe$_{60}$B$_{20}$ grows extremely smooth because of the B content and its amorphous nature. Post-growth annealing at 300°C triggers diffusion of the B atoms into the Ta buffer layer, and CoFe crystallization is initiated from the MgO interface. The MgO|CoFe interface is known to exhibit an exceptionally high out-of-plane magnetic anisotropy after crystallization.[83] Because of the lack of boron after annealing, the CoFeB films will be denoted as

CoFe in the following. In terms of applications, thin films of CoFeB have been proven very useful in magnetic tunnel junctions with up to 500% tunnel-magnetoresistance ratio[42] and for the generation of skyrmion bubbles[43]. They allow for efficient spin-to-charge-current conversion in double-layer systems[44] and low Gilbert damping.[45] Consequently, CoFeB is one of the leading materials for spintronic applications such as the spin-transfer-torque magnetic random-access memory and magnetic read heads and sensors.[46]

Ferrimagnetic $Gd_{27}Fe_{73}$ and $DyCo_5$ alloys were grown by magnetron sputtering with the following stacking sequence: Ta(3 nm)|X(20 nm)|Ta(5 nm)||$Si_3N_4$(150 nm) with X being either $Gd_{27}Fe_{73}$ or $DyCo_5$ (see Supporting Information S3). Both systems have a remanence magnetization state close to its saturation magnetization (see Figure 1 and Ref. 47). The chosen composition of $Gd_{27}Fe_{73}$ and $DyCo_5$ ensures an out-of-plane magnetic anisotropy and a magnetization compensation temperature that is far above[47] ($DyCo_5$) or below[48] ($Gd_{27}Fe_{73}$) the measurement temperature (300 K). The coercive magnetic fields can be reached with moderate external magnetic-field strengths, which are limited to about 150 mT in our experiment. In the following, we refer to $Gd_{27}Fe_{73}$ as GdFe for brevity. $DyCo_5$ and GdFe belong to the class of ferrimagnetic compounds consisting of rare-earth (RE) and transition-metal (TM) elements. They are interesting for spintronic applications because they exhibit a large SOI, highly tunable magnetic properties and large magnetooptical effects.[49] Another intriguing phenomenon discovered recently on these RE-TM ferrimagnetic alloys is all-optical ultrafast magnetization switching,[50,51] which bears a large potential for magnetic recording.

## 2.3 Sample characterization

To characterize the magnetic properties of the samples, we measure the Faraday rotation with a continuous-wave laser diode (wavelength of 628 nm) as a function of an external magnetic oriented normal to the sample plane under an angle of incidence of 45°. The measured square-like hysteresis curve confirms that the magnetic easy axis is out of the sample plane (see **Figure 2b** for $DyCo_5$). DC magneto-transport measurements are conducted on metallic layers that are patterned into Hall-bar structures (Supporting Information S5), the results of which are discussed below). For a complete characterization of the sample's THz conductivity tensor, we also measure the THz transmission of our samples. As a reference, we use samples without metal films as well as only dry air.

## 2.4 Measurement procedure

The sample magnetization **M** is saturated by an external magnetic field (up to $\pm 180$ mT) and typically switched every 10 s. THz waveforms are averaged over about 1000 cycles. Measurements are performed in remanence, except for the GdFe sample, where an external field of about ±50 mT is applied owing to the slight non-square like hysteresis curve (Supporting Information S1-S2). We emphasize that we employ a 45°-analyzer configuration, which does not require rotation of the THz polarizer. Importantly, this approach is different from the often-used nearly-crossed polarizer-analyzer configurations used in conventional ellipsometry measurements.

By not moving any THz optics, we minimize systematic errors arising from, for instance, inhomogeneities of the moving polarizer, which are more pronounced at higher THz frequencies. We cover the entire frequency range from 1 to 40 THz by generating THz radiation subsequently with a spintronic THz emitter[12] (TeraSpinTec GmbH) and a GaSe nonlinear optical crystal. All measurements are conducted at room temperature in a dry $N_2$ atmosphere to avoid THz absorption by water vapor.

## 3. Results

### 3.1 Raw data

Figure 2a displays the electro-optic signal of THz pulses obtained after transmission through a $DyCo_5$ sample for opposite sample magnetizations. At first glance, that is, on the large scale, the THz signals for magnetization $+\mathbf{M}$ (red solid curve) and $-\mathbf{M}$ (black dashed curve) agree almost perfectly. There is, however, a small signal change for opposite magnetizations, which only becomes apparent by magnifying the signal in the vicinity of the maximum at $t = 103$ fs (see inset of Fig. 2a). The magnification reveals that switching between the two magnetizations induces signal changes on the order of 1%.

To evaluate these data, we assume the measured signal $S = S_0 + \Delta S$ to be a sum of signals $S_0$ and $\Delta S$ which are, respectively, independent of $\mathbf{M}$ and linear in $\mathbf{M}$. Effects of higher order in $\mathbf{M}$ are neglected. In this approximation, we obtain

$$S_0 = \frac{S(+\mathbf{M}) + S(-\mathbf{M})}{2} \text{ and } \Delta S = \frac{S(+\mathbf{M}) - S(-\mathbf{M})}{2}. \tag{1}$$

By applying this procedure to the waveforms of Figure 2a, we find the magnetization-dependent signal is on the order of $\Delta S/S_0 \sim 2\%$. In addition, $S(+\mathbf{M})$ and $S(-\mathbf{M})$ are almost in phase (Inset of Figure 2a), indicating that the transmitted THz pulse is still linearly polarized.

As a check, we also perform a reference measurement of the bare substrate (150 nm thick $Si_3N_4$ membrane), which does not result in any detectable signal odd in the sample magnetization (Figure S1). To further verify the magnetic origin of the signal, we measure the complete THz waveform as a function of the external magnetic field $\mathbf{B}_{\text{ext}}$. Figure 2b shows the root mean square (RMS) of these THz waveforms vs. $\mathbf{B}_{\text{ext}}$. We find that the THz hysteresis curve follows the optical Faraday rotation hysteresis curve. Therefore, $\Delta S$ indeed scales with the sample magnetization.

### 3.2 Conductivity-tensor extraction

The signals shown in Figure 2 still depend on sample-extrinsic factors such as the spectrum of the incident THz pulse and the sample substrate. Significantly more information is provided by extracting the conductivity tensor of the metallic magnet from our data. For the analysis, it is sufficient to restrict oneself to the *x-y*-plane (Figure 1) because all currents flow in the sample plane. In an isotropic magnetically ordered solid with magnetization $\mathbf{M} \| \mathbf{u}_z$, the current $\mathbf{j}(\omega, z)$ driven by an electric field $\mathbf{E}(\omega, z)$ is at frequency $\omega/2\pi$ and position $z$ given by

$$\mathbf{j} = \underline{\sigma}\mathbf{E} = \begin{pmatrix} \sigma_{xx} & -\sigma_{yx} \\ \sigma_{yx} & \sigma_{xx} \end{pmatrix} \mathbf{E}. \tag{2}$$

Here, $\mathbf{E}$ is the electric field inside the sample, and $\sigma_{xx}$ and $\sigma_{yx}$ denote the diagonal and off-diagonal conductivity, respectively. Note that the Onsager relations and Equation (2) imply that $\sigma_{yx}(-\mathbf{M}) = \sigma_{xy}(\mathbf{M}) = -\sigma_{yx}(\mathbf{M})$.

The connection to our experiment (Figure 1b) is provided by the Fresnel transmission matrix $\underline{t} = (t_{ij})$, which relates the incident and transmitted electric fields by $\mathbf{E}_{\text{out}} = \underline{t}\mathbf{E}_{\text{inc}}$. Here, the indices $i, j$ equal s or p, which correspond to the $x$ and $y$ axes in a normal-incidence geometry, respectively. In our setup, we do not measure the electric field directly but an electro-optic signal $S$, which is in the Fourier domain related to the electric field by multiplication with a frequency-dependent setup transfer function. The detector is equally sensitive to s- and p-polarized THz fields because the angle of the THz polarizer behind the sample is set to 45°. By acquiring $S$ for opposite magnetizations $\pm\mathbf{M}$, we obtain the nonmagnetic signal $S_0(\omega)$ and the magnetic signal $\Delta S(\omega)$. By taking the ratio, the setup transfer function drops out and we obtain

$$\frac{\Delta S}{S_0} = \frac{\Delta E}{E_0} = \frac{t_{\text{sp}}}{t_{\text{pp}}}. \tag{3}$$

The coefficient $t_{\text{pp}}$ is obtained by an additional reference measurement without sample and using the optical constants for $Si_3N_4$ as indicated in Table S1.

To approximately determine the information contained in Equation (3), we apply the thin-film approximation.[52] The two Fresnel coefficients for normal incidence become

$$t_{\text{pp}}(\omega) = \frac{2n_1(\omega)}{n_1(\omega) + n_2(\omega) + Z_0 G_{xx}(\omega)} \tag{4}$$

and

$$t_{\text{sp}}(\omega) = \frac{t_{\text{pp}}^2(\omega)}{2n_1(\omega)} Z_0 G_{yx}(\omega) \tag{5}$$

where the sheet conductance $G_{ij}$ is given by $G_{ij}(\omega) = \int_0^d dz\, \sigma_{ij}(z, \omega)$, and $Z_0 \approx 377\,\Omega$ is the vacuum impedance. In the special case of frequency-independent optical constants, normal incidence, a homogeneous layer of thickness $d$ between two air half spaces (refractive index $n_1 = n_2 = 1$) and in the limit $Z_0 G_{xx} \ll 1$, one finds $\Delta S/S_0 \approx Z_0 G_{yx}/2$, which is directly proportional to $\sigma_{yx}$. The diagonal sheet conductance $G_{xx} = G_{xx}^{\text{FM}} + G_{xx}^{Si_3N_4} + G_{xx}^{\text{Ta}}$ contains contributions from the $Si_3N_4$ membrane and from the Ta seed and capping layers. They are measured on individual $Si_3N_4$ and $Si_3N_4|$Ta samples, respectively (Supporting Information S3).

Note that the thin-film approximation is valid as long as the metal thickness is much smaller than the THz wavelength and the penetration depth inside the material and at the given THz frequency.[12] In metals, the two above characteristic lengths are on the order of 1 μm and 100 nm at 1 THz.[53]

Thus, for the highest THz frequencies, the thin-film approximation becomes inaccurate for metal films with thicknesses in the 10 nm range. Therefore, we employ a more exact transfer-matrix approach instead to extract the in-plane conductivity tensor $\underline{\sigma}$, which also accounts for the 45° angle of incidence of our experiment (Supporting Information S4). **Figures 3 and 4** show the central result of this procedure: the complex-valued conductivity tensor $\underline{\sigma}$ of all studied materials over more than 5 octaves from 1 to 40 THz.

### 3.3. Diagonal conductivity $\sigma_{xx}$

We start with considering the extracted diagonal conductivities $\sigma_{xx}$ (Figure 3). Note that for DyCo$_5$, we find a good match between DC and THz conductivity. No DC measurements were performed on the other materials.

The frequency-dependence of the conductivity of metals often obeys the Drude formula.[54] It can be derived from the Boltzmann transport equation in the relaxation-time approximation, which considers the conduction electrons as classical particles scattering at the electronic velocity relaxation rate $\Gamma$. For all materials, we observe a typical Drude-like behavior, that is, a monotonically decreasing Re $\sigma_{xx}$ with increasing frequency. By fitting the Drude formula

$$\sigma^{\text{Drude}}(\omega) = \frac{\sigma_{\text{DC}}}{1 - i\omega/\Gamma} \quad (6)$$

to our data, we obtain the DC conductivity $\sigma_{\text{DC}}$ and $\Gamma$ (**Table 1**).

For the extracted average DC conductivity of CoFe(1 nm)|Ta(8 nm), we find good agreement with the value ($5 \times 10^5$ S m$^{-1}$) reported in refs. 55 and 56 for pure CoFe. The agreement indicates that the two materials, CoFe and Ta, have a similar conductivity, as also reported previously.[54] To the best of our knowledge, for DyCo$_5$, only one measurement on a much thicker film exists.[57] The conductivity ($2.8 \times 10^6$ S m$^{-1}$) is one order of magnitude larger than our result, likely because interface scattering makes a smaller relative contribution to electron scattering in thicker films. However, we find good agreement with the conductivity reported for DyCo$_3$ ($3.3 \times 10^5$ S m$^{-1}$), the stoichiometry of which is, however, different from our sample.[58] In the case of GdFe, our measured conductivity matches the reported value of $5.0 \times 10^5$ S m$^{-1}$.[59, 60]

The large current-relaxation rate $\Gamma$ of more than 100 THz implies that the mean time $1/\Gamma$ between subsequent scattering events amounts to just a few femtoseconds. This observation agrees with previous thin-film studies,[61] which assigned the large scattering rates to strong disorder due, for instance, to small grain sizes and significant interface roughness[62] as well as enhanced electron-phonon scattering in disordered alloys.[63,64]

### 3.4 Off-diagonal conductivity $\sigma_{yx}$

Figure 4 shows the anomalous Hall conductivity (AHC) extracted from 1 to 40 THz. We note that a small paramagnetic contribution to the signal due to the ordinary Hall effect and slow drifts of the setup may lead to systematic errors, resulting in a nonvanishing value of Im $\sigma_{yx}(\omega = 0)$. Accordingly, the experimental uncertainties are estimated by deviations from the condition

Im $\sigma_{yx}(\omega = 0) = 0$. We emphasize that we again find good agreement between DC and THz measurements. The slightly smaller DC value may originate from different substrate thicknesses (0.15 vs 500 μm, see Supporting Information S5). The good match between DC and THz measurements at the lowest frequencies is not unexpected in view of a variety of recent THz spintronic studies.[12,65,66,67] In comparison to reported values, we find good agreement for GdFe ($2.5 \times 10^4$ S m$^{-1}$ in Ref. 57). For DyCo$_5$, the values reported in Ref. 55 for much thicker films are 3 times larger ($3.3 \times 10^4$ S m$^{-1}$). For CoFe, our measured AHC agrees well with *ab-initio* calculation of the intrinsic AHC ($3.3 \times 10^4$ S m$^{-1}$ in Ref. 68).

We find that Re $\sigma_{yx}$ only slightly changes toward low THz frequencies (Figure 4), whereas Im $\sigma_{yx}$ exhibits an approximately linear decrease toward low frequencies. This behavior is consistent with the fact that Re $\sigma_{yx}$ is even with respect to $\omega$, whereas Im $\sigma_{yx}$ is odd, because the conductivity is a real-valued quantity in the time domain. Notably, the overall spectral behavior of $\sigma_{yx}$ is qualitatively analogous to $\sigma_{xx}$.

## 4. Discussion

To discuss the frequency dependence of the measured $\sigma_{yx}$ (Figure 4), we first review the microscopic mechanisms that contribute to the AHE: (i) The intrinsic contribution, which is already present in a perfect crystal, and the extrinsic mechanisms (ii) skew scattering and (iii) side jump.[4,5,6]

### 4.1 Intrinsic AHE contribution

In many theoretical considerations in the DC limit[5,6], the intrinsic contribution (i) is often discussed in terms of the anomalous velocity, which is a velocity component perpendicular to the driving electric field. The anomalous velocity scales linearly with the instantaneous value of the driving electric field and the real-valued Berry curvature, which results in a conductivity component $\sigma_{yx}$ that is independent of frequency along with Im $\sigma_{yx} = 0$. This notion does not agree with our observations: For GdFe, for instance, $\sigma_{yx}$ changes strongly with frequency (Figure 4). In particular, Im $\sigma_{yx}$ increases with frequency.

We note that the concept of the anomalous velocity is only valid at sufficiently low frequencies of the driving field[5]. At arbitrary frequencies, the AHC can be calculated within Kubo linear-response theory using[69]

$$\sigma_{yx}^{\text{calc}}(\omega) = \frac{ie^2\hbar}{m^2 V} \sum_{\mathbf{k},n,n'} \frac{f(\epsilon_{\mathbf{k}n}) - f(\epsilon_{\mathbf{k}n'})}{\epsilon_{\mathbf{k}n} - \epsilon_{\mathbf{k}n'}} \frac{\langle \mathbf{k}n'|\hat{p}_y|\mathbf{k}n\rangle \langle \mathbf{k}n|\hat{p}_x|\mathbf{k}n'\rangle}{\epsilon_{\mathbf{k}n'} - \epsilon_{\mathbf{k}n} + \hbar\omega + i\hbar\gamma}. \tag{7}$$

Here, the involved quantities are the matrix elements of the momentum operator $p$, the Bloch band energies $\epsilon_{\mathbf{k}n}$, initial and final Bloch states $|\mathbf{k}n\rangle$ and $|\mathbf{k}n'\rangle$, the system volume $V$, the electron mass $m$, and the Fermi-Dirac function $f(\epsilon_{\mathbf{k}n})$. In the limit $\omega \to 0$ and $\gamma \to 0$, Equation (7) leads to the frequently used expression for the DC AHC in terms of the Berry curvature.[5]

Our calculations, differently from those presented in Ref. 70, also introduce the effect of the Bloch electrons' lifetimes quantified by the inverse lifetime $\gamma$ of the state. We note that even though this spectral broadening is introduced phenomenologically, a suitable range of values is known for metals.[64] For transitions near the Fermi energy, $\gamma$ is expected to be of the same order as the relaxation rate $\Gamma$ of the Drude formula (see Equation (6)).

The summation over the band indices $n, n'$ and wavevectors **k** accounts for all allowed transitions. The Bloch states, band energies and momentum matrix elements are computed using a relativistic density-functional theory implementation.[71] Note that the contribution from intraband transitions ($n = n'$) in Equation (7) is zero, in contrast to the diagonal tensor element $\sigma_{xx}(\omega)$, where the intraband contribution leads to a Drude-like conductivity. The $\sigma_{xx}(\omega)$ can in principle be computed *ab-initio*, but it strongly depends on the details of the sample quality and is, thus, not done here.

The calculated $\sigma_{yx}^{\text{calc}}$ are shown in **Figure 5** from 0 to 100 THz for various choices of $\gamma$. For $\hbar\gamma = 137$ meV, we find good agreement with the complex-valued measured $\sigma_{yx}$ (see Figure 4 for CoFe). This observation is consistent with the expectation $\gamma = \Gamma/2$, which can be derived by comparison of the Drude and the Kubo formula (Equation (6) and (7)). The substantial frequency dependence of $\sigma_{yx}^{\text{calc}}$ arises from resonant interband transitions for which $\epsilon_{kn'} - \epsilon_{kn} + \hbar\omega \approx 0$ in Equation (7). These spectral features are particularly pronounced for smaller values of $\gamma$ (Figure 5). Similar observations were reported for $SrRuO_3$ at temperatures at about 10 K and frequencies of around 1 THz (Ref. 26).

In contrast, our measured frequency dependence is featureless (Figure 4). Comparison with the *ab-initio* computed AHC suggests that the flat frequency response of $\sigma_{yx}$ arises from the large scattering rate of the electrons. The large value of $\gamma$ leads to a significant broadening of electronic transitions underlying the AHE and so smears out sharp spectral features of $\sigma_{yx}$ observed for the smallest broadening in Figure 5.

### 4.2 External vs proper conductivity

We note that the conductivity calculated by the Kubo formalism (see Equation (7)) is called external (likewise direct or full) conductivity because it relates the external (incident) electric field to the current driven inside the sample. In contrast, the measured conductivity refers to the total electric field **E** (incident plus reaction field) inside the sample (Equation (2)).[72] It is called proper conductivity.

To account for this effect (Supporting Information S7), we multiply the calculated external off-diagonal conductivity $\sigma_{yx}^{\text{calc}}$ by $(1 + Z_0\sigma_{xx}d/2)^2$, as shown in Figure 4 (dark and light green curves) and in Figure S4 for other broadenings. While the agreement of experiment and theory for Im $\sigma_{yx}(\omega)$ remains effectively the same, it has improved noticeably for Re $\sigma_{yx}(\omega)$. We conclude that the intrinsic AHE mechanism can well explain the measured proper $\sigma_{yx}(\omega)$ of CoFe, both in terms of magnitude and frequency dependence, provided a sufficient lifetime broadening is introduced. Even better agreement is obtained when one accounts for the fact that the calculated conductivity (Equation (7)) is the external conductivity.

The observed discrepancy between theory and experiment of only about 20% may arise from extrinsic contributions to the AHE, the different stoichiometries employed for CoFe in the experiment and in the calculation (see Supplementary Information S6) as well as substrate-induced stress in the thin metal films, all of which are neglected in the calculations.

### 4.3 Extrinsic AHE contributions

We note that in Equation (7), Bloch states are assumed to be the single-particle states for the electrons of the unperturbed system. The extrinsic AHE contributions skew scattering (ii) and side-jump[5,6] (iii) are neglected. In the following, we estimate the strength of (ii) and (iii).

For the skew-scattering contribution (ii), we employ a simple Boltzmann-equation-type model (Supporting Information S8). The resulting dependence on frequency (but not on $\Gamma$) is identical to that of the ordinary Hall effect (see Equation (21)): We find the Drude-type relationship of Equation (6) for $\sigma_{xx}$, whereas $\sigma_{yx}^{sk}(\omega) \propto (1 - i\omega/\Gamma)^{-2} \approx (1 - 2i\omega/\Gamma)^{-1}$. Therefore, the normalized imaginary part $\text{Im}\,\sigma_{yx}^{sk}(\omega)/\sigma_{xy}(0)$ of the skew-scattering component should increase with slope $(2/\Gamma)\sigma_{yx}^{sk}(0)/\sigma_{yx}(0)$.

To obtain an upper limit of the skew-scattering contribution for CoFe, we assume that the measured slope of $\text{Im}\,\sigma_{yx}(\omega)$ exclusively arises from skew scattering. We obtain $\sigma_{yx}^{sk}(0)/\sigma_{yx}(0) < 0.3$, that is, skew scattering would at most contribute 30% to the AHC. In this estimate, we assumed similar values of $\Gamma$ for CoFe and Ta in the CoFe(1 nm)|Ta(8 nm) bilayer as justified by measurements of $\sigma_{xx}(\omega)$ in pure Ta films (Figure S2 and Table 1). This result is in line with previous notions[5] that the low conductivity of our sample places it in the bad-metal regime. There, skew-scattering is known to make a rather negligible contribution to $\sigma_{yx}$, and the intrinsic mechanism dominates.

Concerning the side-jump component (iii), we note that its contribution is difficult to discriminate against the intrinsic contribution.[5] However, side-jump is typically one order of magnitude weaker than the intrinsic contribution.[71]

To summarize, the good agreement of our calculations based on Equation (7) with the measured data (Figure 4) and the estimated small magnitude of less than 25% of the skew-scattering contribution strongly suggests that the intrinsic mechanism dominates $\sigma_{yx}$ at least for CoFe. This conclusion appears reasonable because the intrinsic contribution is enhanced by the large SOI of d-electrons around the Fermi energy, as in our samples. Thus, the large quasiparticle scattering rate in our samples has two important consequences: It suppresses the extrinsic skew-scattering contribution, and it smears out any spectral feature in $\sigma_{yx}$.

### 4.4 Anomalous Hall angle

We finally consider the anomalous Hall angle (AHA), which is defined as $\Theta_{AH}(\omega) = \sigma_{yx}(\omega)/\sigma_{xx}(\omega) = \Delta j(\omega)/j_0(\omega)$ and displayed for DyCo$_5$ and GdFe in **Figure 6**. The $\Theta_{AH}$ of CoFe could not be determined because its $\sigma_{xx}$ could not be separated from that of Ta in the CoFe|Ta sample. We find a largely frequency-independent $\Theta_{AH}$ from DC to 40 THz with frequency-averaged values of 2.6% for DyCo$_5$ and 2.9% for GdFe (Figure 6). The THz AHA of DyCo$_5$ approximately agrees with its measured DC value of 2.0%. Note that the driving and the AHE-induced electric field components, i.e. $\Delta E$ and $E_0$, are in phase (Figure 2a), already indicating a real-valued AHA for DyCo$_5$. Indeed, in the electrostatic limit, one has $\Delta E/E_0 = \text{Re}\,\Theta_{AH} \approx 0.02$ (Supporting Information S5), consistent with the raw data in Figure 2a. We find good agreement between our measured AHA values and reported values for GdFe (2.5 % in Ref. 73 and 4 % in Ref. 74). We are not aware of any reported $\Theta_{AH}$ for DyCo$_5$.

For DyCo$_5$, Im $\Theta_{AH}$ is relatively small for all frequencies, whereas for GdFe, Im $\Theta_{AH}$ monotonically increases with $\omega$ up to 30% of Re $\Theta_{AH}$ at 40 THz. The positive slope of Im $\Theta_{AH}(\omega)$ implies that the AHE-induced current $\Delta j$ lags behind the primary current $j_0$ (Figure 1) by the group delay $\partial \arg \Theta_{AH}(\omega)/\partial \omega \approx 1.2$ fs. Therefore, in our experimental frequency range, the bandwidth of $\sigma_{yx}$ is smaller than the bandwidth of $\sigma_{xx}$, which is on the order of $\Gamma$. A possible reason for the reduced bandwidth of $\sigma_{yx}$ is suggested by Equation (7): Bloch states with strong SOI exist only in a limited energy range significantly smaller than $\hbar\Gamma$ around the Fermi energy. Thus, only transitions with frequencies $|\omega|$ significantly below $\Gamma$ contribute to the sum of Equation (7).

## 5. Conclusion

In summary, we developed a technique to measure the AHE in metals continuously from DC to 40 THz, which is a highly relevant spectral window with respect to SOI energy scales. In the studied materials, the AHE is operative even at the highest THz frequencies. For DyCo$_5$, we explicitly confirmed the consistency of our high-frequency with DC measurements. The quantitative agreement with *ab-initio* calculations for CoFe leads us to the conclusion that the intrinsic AHE contribution dominates and that the spectrally flat off-diagonal conductivity originates from the large quasi-particle scattering rates.

Since the intrinsic contribution to the SHE and the AHE share the same physical origin at zero frequency, the Berry curvature,[4] our results strongly suggest that also the intrinsic SHE contribution of metals is largely frequency-independent up to 40 THz. This conclusion agrees with calculations of the SHE conductance of Pt and W, which found a constant value up to about 100 THz.[75] Our study, thus, closes the gap between DC and the THz range for both AHE and SHE.

Future studies based on our methodology will permit even more insight into SOI at THz frequencies. More pronounced spectral features in the AHC are theoretically expected for samples with small broadening. Experimentally, such reduced level broadening can be achieved by either measuring at low temperatures[33,76] or by using samples with fewer impurities. In principle, we see no obstacle to employ our broadband technique at low temperatures. As Kim *et al.*[33] showed, a direct distinction between extrinsic and intrinsic contributions to the AHE conductivity becomes possible if $\Gamma/2\pi$ lies in the experimentally covered frequency window.[21] This case is especially interesting for systems with similar-sized intrinsic and extrinsic effects at DC such as L1$_0$ FePt (Ref. 77). Finally, extending this measurement scheme to nonmagnetic materials by means of spin injection will allow one to all-optically observe the dynamics of the SHE.[17]

**Acknowledgements**

T.S.S. and T.K. would like to express their gratitude to Viktor Platschkowski whose engagement and passion not only greatly facilitated this work but also will be deeply missed. We acknowledge the European Research Council for funding through the projects TERAMAG/Grant No. 681917 and ECOMAGICs/Grant No. 280048, the European Union's Horizon 2020 FET Open program for funding through FEMTOTERABYTE/Grant No. 737709, and ASPIN/Grant No. 766566, the German Research Foundation for funding through the CRC/TRR 227 "Ultrafast spin dynamics" (projects A05, B01, B02 and MF) and the projects MU 1780/8-1 and MU 1780/10-1, the Swedish Research Council (VR), the K. and A. Wallenberg Foundation (Grant No. 2015.0060), the German Ministry for Education and Research (BMBF) through project 05K16BCA Femto-THz-X, and Ministry of Education of the Czech Republic Grant No. LM2018096, LM2018110, and LNSM-LNSpin, Czech Science Foundation Grant No. 19-28375X. The authors acknowledge financial support from the Horizon 2020 Framework Programme of the European Commission under FET-Open Grant No. 863155 (s-Nebula). They also acknowledge computer time received from the Swedish National Infrastructure for Computing (SNIC).

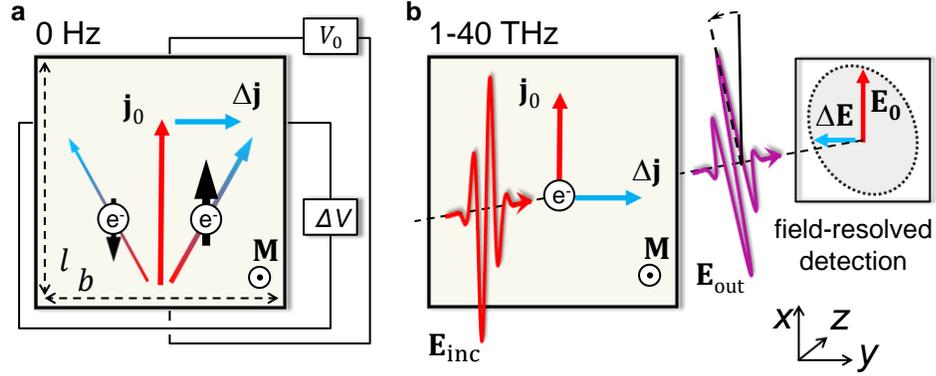

**Figure 1. Schematic of DC and THz anomalous Hall effect measurements. a**, In the DC AHE measurement, an applied voltage $V_0$ drives a DC charge current $\mathbf{j}_0 = j_0 \mathbf{u}_x$ parallel to the $x$ axis in magnetic metallic sample of length $l$ and width $b$. Spin-orbit interaction (SOI) deflects spin-up and spin-down electrons (black arrows) into opposite directions perpendicular to their velocity and to the sample magnetization $\mathbf{M} \parallel \mathbf{u}_z$. The different number of majority (spin-up) and minority (spin-down) electrons causes a perpendicular charge current $\Delta \mathbf{j} = \Delta j \mathbf{u}_y$ that leads to a corresponding voltage $\Delta V$ measured by a voltmeter. **b**, In the all-optical AHE measurement, an incident THz electromagnetic pulse with transient electric field $\mathbf{E}_{\text{inc}} \parallel \mathbf{u}_x$ drives an AC charge current $\mathbf{j}_0 = j_0 \mathbf{u}_x$ with frequencies from 1 to 40 THz in the plane of the magnetic metal along $\mathbf{E}_{\text{inc}}$. SOI induces a transverse current $\Delta \mathbf{j} = \Delta j \mathbf{u}_y$, which emits an additional THz electric field component $\Delta \mathbf{E}$ into the optical far-field. The superposition $\mathbf{E}_0 + \Delta \mathbf{E}$ leads to an elliptically polarized THz wave behind the sample. With an electric-field-sensitive detector, both THz polarization components $\mathbf{E}_0$ and $\Delta \mathbf{E}$ are separately measured with femtosecond time resolution.

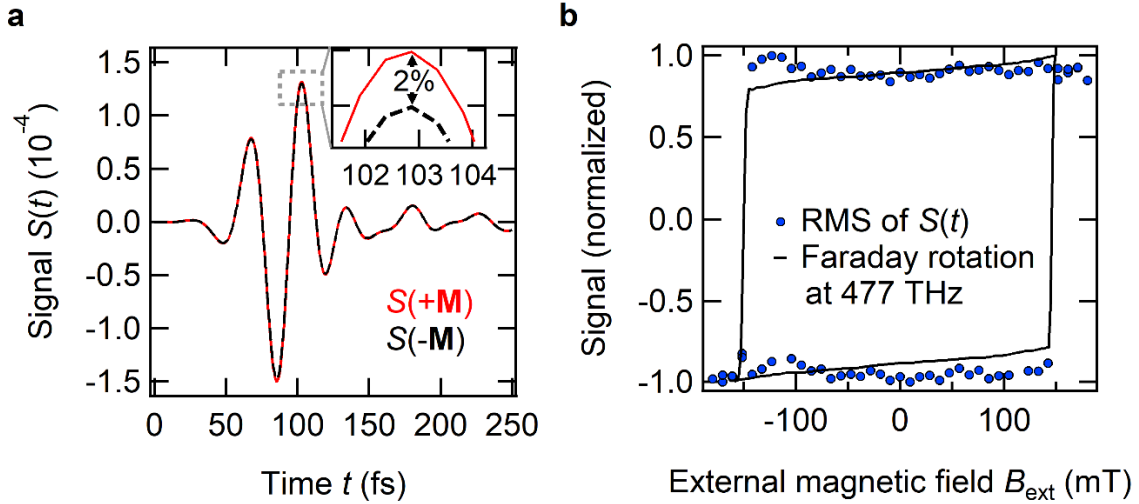

**Figure 2. Raw data of the THz AHE of DyCo$_5$. a**, THz signal of an initially linearly polarized THz pulse after passing through an out-of-plane magnetized DyCo$_5$ sample. The anomalous Hall effect induces a new perpendicular polarization component, depending on the orientation of the sample magnetization (red solid and black dashed curve). **Inset**: Magnification reveals a signal change on the order of 2 %, suggesting an anomalous Hall angle of similar magnitude. **b**, Faraday rotation hysteresis curve at optical (628 nm ≙ 477 THz, black solid line) and terahertz (10 to 40 THz, blue circles) frequencies (root mean square (RMS) of the THz waveform $\Delta S(t)$ (Equation (1)) multiplied by its polarity).

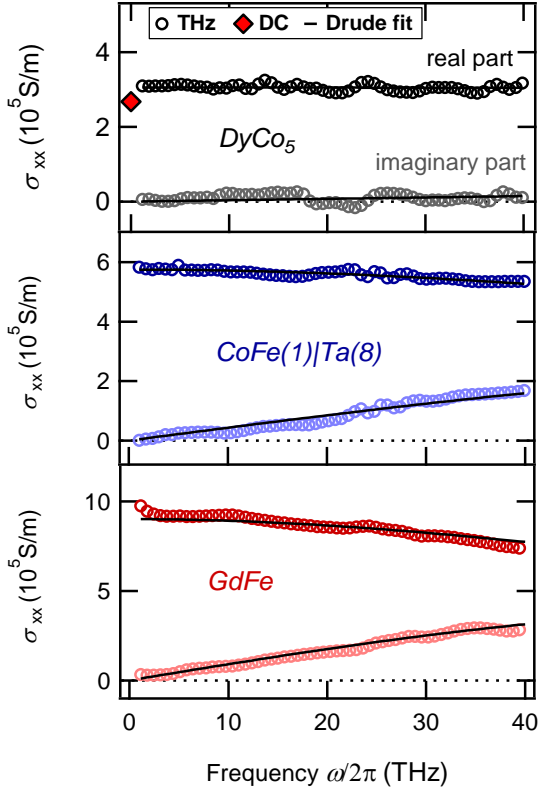 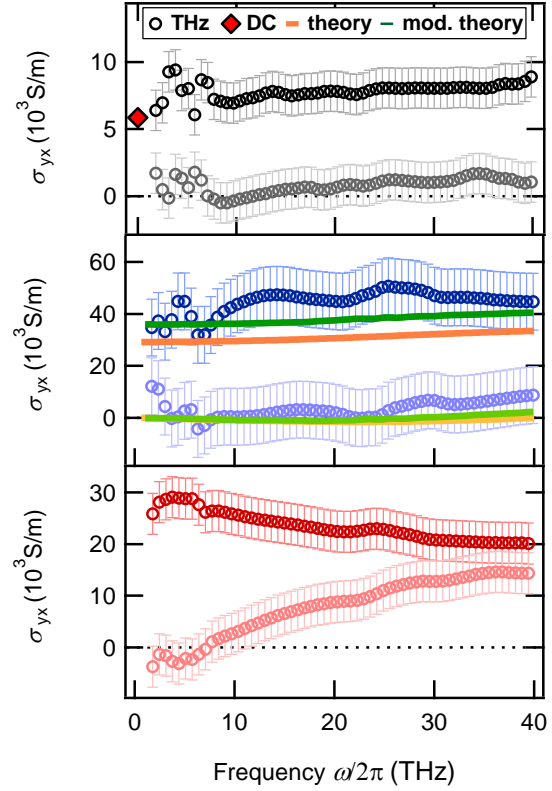

**Figure 3. Measured complex-valued diagonal conductivity of DyCo₅, CoFe and GdFe from DC to 40 THz.** Diagonal conductivities $\sigma_{xx}$ measured in the THz frequency range (real part: dark circles, imaginary part: light circles) and at DC (red diamond symbol). For the CoFe sample, only the average conductivity of the CoFe(1 nm)|Ta(8 nm) stack was extracted (see Supporting Information S4). Fits (solid lines) were obtained using the Drude model (see Equation (6) and Table 1).

**Figure 4. Measured and *ab-initio*-calculated complex-valued anomalous Hall conductivity of DyCo₅, CoFe and GdFe from DC to 40 THz.** Off-diagonal conductivities $\sigma_{yx}$ measured in the THz frequency range (real part: dark circles, imaginary part: light circles) and at DC (red diamond symbol). Results of *ab-initio* calculations are shown by solid lines for a broadening of $\hbar\gamma = 137$ meV (see Equation (7)). While the orange/yellow lines are the $\sigma_{yx}^{\text{calc}}$ with respect to the external perturbing field (Equation (8) in the supporting information), the green/light-green lines are the conductivity with respect to the external and reaction field (Equation (9) in the supporting information). Experimental errors

are estimated from deviations from the condition Im $\sigma_{yx} = 0$ at $\omega = 0$.

| **Material** | $\sigma_{DC}(10^5$ S/m) | $\Gamma/2\pi$ (THz) |
|---|---|---|
| DyCo$_5$ | 3.1 | 830 |
| CoFe(1 nm) \|Ta(8 nm) | 5.8 | 130 |
| GdFe | 9.0 | 100 |
| Ta (see Fig. S2) | 5.7 | 130 |

**Table 1.** Drude model fit parameters for the experimental diagonal-conductivity data in Figure 3a.

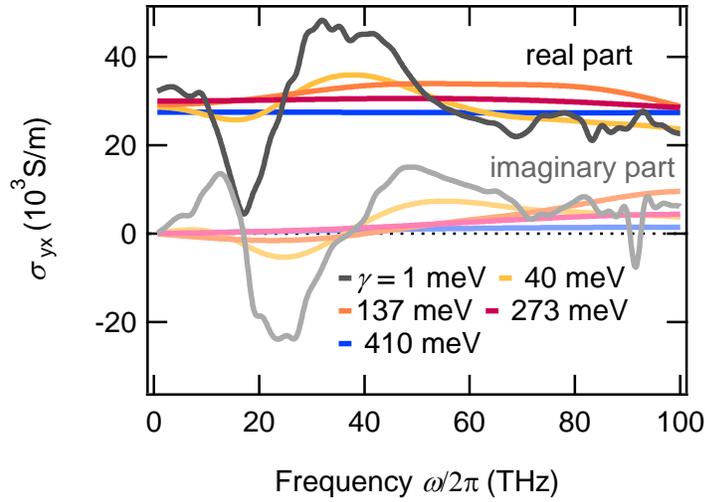

**Figure 5.** *Ab-initio*-**calculated anomalous Hall conductivity**. Theoretical off-diagonal conductivity $\sigma_{yx}^{calc}$ of CoFe calculated for different broadenings according to Equation (7). The light-colored lines are the respective imaginary parts.

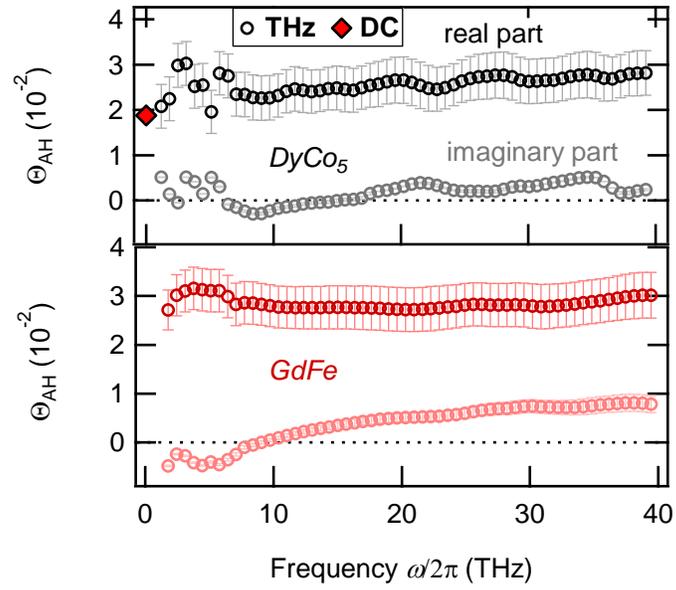

**Figure 6. Measured complex-valued THz anomalous Hall angles of DyCo$_5$ and GdFe.** Anomalous Hall angles $\Theta_{AH}$ measured in the THz frequency range (real part: dark circles, imaginary part: light circles) and at DC (red diamond symbol). Experimental errors are estimated from the uncertainty in $\sigma_{yx}$ (see Figure 4).

# Supporting Information

**Frequency-independent terahertz anomalous Hall effect in DyCo$_5$, Co$_{32}$Fe$_{68}$ and Gd$_{27}$Fe$_{73}$ thin films from DC to 40 THz**

*Tom S. Seifert, Ulrike Martens, Florin Radu, Mirkow Ribow, Marco Beritta, Lukas Nádvorník, Ronald Starke, Tomas Jungwirth, Martin Wolf, Ilie Radu, Markus Münzenberg, Peter M. Oppeneer, Georg Woltersdorf, Tobias Kampfrath*

## S1 Technical details of the THz spectrometer

We use 80% of the output of a Ti:sapphire laser oscillator (pulse duration 10 fs, center wavelength 800 nm, pulse energy 2.5 nJ, repetition rate 80 MHz) to generate THz pulses by difference-frequency mixing in a nonlinear-optical material, either a GaSe crystal (thickness of 90 µm) or a spintronic THz emitter Co$_{40}$Fe$_{40}$B$_{20}$(3 nm)|Pt(2 nm) (TeraSpinTec GmbH).[1,2] First, the THz pulse traverses a wire grid polarizer (WGP), making it p-polarized (residual ellipticity < $10^{-2}$ with respect to field amplitude). Subsequently, a parabolic mirror focuses the THz beam onto the sample under an angle of incidence of 45°. Behind the sample, the beam is collimated by a second parabolic mirror. Afterwards, a second WGP set to an angle of 45° with respect to the table plane makes the detector equally sensitive to s- and p-polarized THz light. For the detection, the THz beam is focused onto an electro-optic crystal (ZnTe(110) (thickness of 10 µm) or GaP(110) (250 µm)) where its transient electric field is sampled using a co-propagating 10 fs laser pulse (energy of 0.5 nJ).[3] By using various emitter and detector combinations, we are able to cover the entire range from 1 to 40 THz.

To increase the signal-to-noise ratio, we combine lock-in detection of the signal $S$ and rapid scanning of $S$ vs $t$. For this purpose, we modulate the amplitude of the THz beam with a mechanical chopper (frequency of 30 kHz) and vary the delay $t$ at a frequency of 25 Hz with a so-called shaker (APE GmbH, Berlin/Germany). The result is a shot-noise limited detection of the electro-optic signal $S$.[4]

## S2 Reference measurement

To exclude any magnetic signals that are not coming from the actual samples, we performed a reference measurement using a Si$_3$N$_4$ membrane without any metal film. Figure S1 shows that there is no signal within the detection sensitivity.

## S3 Sample details

***CoFe.*** The Co$_{20}$Fe$_{60}$B$_{20}$ film with the layer stacking MgO(2 nm)|Co$_{20}$Fe$_{60}$B$_{20}$(1 nm)|Ta(8 nm)||Si$_3$N$_4$(150 nm) from the Greifswald group was prepared by magnetron sputtering. The MgO capping and tantalum metal buffer layers were grown by electron-beam evaporation under ultrahigh-vacuum conditions using in situ transfer. The base pressure in the vacuum preparation chambers was of 5 × 10−10 mbar. Similar to the preparation of MgO|CoFeB|Ta layers for magnetic tunnel junctions, the samples were post

growth annealed at 300°C for 1h. This annealing allows a diffusion of the boron into the tantalum layer that acts as a boron sink and crystallization of the previously amorphous grown films. This process is called solid-state epitaxy.

A high-resolution transmission-electron-microscopy study from some of the authors can be found in Ref. 5, which shows in detail the process of the solid-state epitaxy and the crystallite interface matching in the (001) growth direction of Fe and MgO. The tunneling magneto resistance ratio in typical junctions $Co_{20}Fe_{60}B_{20}|MgO|Co_{20}Fe_{60}B_{20}|Ta$ reaches 150-270%.[5] A general composition analysis of the films yields a Co-Fe ratio of 1/2.1, and characterization by transmission electron microscopy revealed smooth $Co_{20}Fe_{60}B_{20}$ film surfaces below the atomic monolayer limit. The samples exhibited strong out-of-plane magnetic anisotropy, originating from the MgO|CoFe interface, and a nearly rectangular hysteresis curve with a coercive field well below 2-3 mT.[6]

***$Gd_{27}Fe_{73}$ and $DyCo_5$.*** The $DyCo_5$(20 nm) and $Gd_{27}Fe_{73}$(20 nm) thin-films were grown on $Si_3N_4$ substrates (150 nm thick window supported by a 500 μm thick frame) by magnetron sputtering using the MAGSSY deposition chamber at Helmholtz-Zentrum Berlin. As buffer and capping layers, we used Ta thin-films of 5 nm and 3 nm thickness, respectively. The layers have been deposited in an ultraclean Argon atmosphere of $1.5 \cdot 10^{-3}$ mbar with a base pressure below $5 \cdot 10^{-9}$ mbar. To avoid inter-diffusion, the deposition temperature was kept at 300 K. The stoichiometry of the ferrimagnetic alloys was controlled by varying the deposition rate of separate chemical elements in a co-evaporation scheme. The used $Si_3N_4$ membranes promote an amorphous or a polycrystalline growth of the samples depending on the relative elemental stoichiometry of the deposited alloy.

Generally, the off-stoichiometric phases of the $Gd_xFe_{1-x}$ alloys are structurally amorphous except for the few existing stoichiometric phases, the so-called Laves phases, for instance $GdFe_2$, $Gd_6Fe_{23}$ and $Gd_2Fe_{17}$. Unless annealed to very high temperatures (up to 800 °C, see e.g. Ref. 7), at which re-crystallization occurs, the thin-film GdFe samples remain amorphous, as it is the case for the $Gd_{27}Fe_{73}$ sample studied here. The amorphous growth and elemental stoichiometry distribution of $Gd_{24}(FeCo)_{76}$ samples deposited on $Si_3N_4$ membranes (similar to those studied here) have been reported in Ref. 8 using transmission electron microscopy and energy dispersive X-ray analysis techniques.

The $DyCo_5$ sample is a stoichiometric phase of $Dy_xCo_{1-x}$ alloys and was shown to be crystalline when grown on single-crystalline substrates such as MgO(110) or/and $Al_2O_3$(11-20). Based on our extensive experience in growing $DyCo_5$ and other $Dy_xCo_{1-x}$ alloys on various substrates[9,10] and on the fact that $DyCo_5$ is a stoichiometric phase, it is reasonable to assume that the $DyCo_5$ films grown on $Si_3N_4$ membranes are polycrystalline. However, the precise crystalline state of the sample has no impact on the conclusions of this work.

Thus, $Gd_{27}Fe_{73}$ is amorphous while $DyCo_5$ (a stoichiometric phase of the Dy-Co alloys) is polycrystalline.

**S4 Extraction of the in-plane conductivity tensor $\underline{\sigma}$**

As imposed by Equation (1) in the main text, an incident electric field $\mathbf{E}_{\text{inc}} = E_0 \mathbf{u}_x$ causes a current $\mathbf{j} = j_0 \mathbf{u}_x + \Delta j \mathbf{u}_y$ inside the ferromagnetic sample. Through electric-dipole radiation, the transverse current $\Delta j$ leads to an outgoing elliptically polarized transmitted THz electric field $\mathbf{E}_{\text{out}}$ behind the sample.

Note that in our time-domain experiment, we measure an electrooptic signal $S(t)$, which is related to an electric-field component $E(t)$ through a convolution with a setup transfer function $h(t)$ that quantifies the propagation of the THz pulse to the detector and its measurement by electrooptic sampling. This convolution turns into a simple multiplication upon Fourier transformation of the signals $S(t)$.

The connection to the experiment is provided by the Fresnel transmission matrix $\underline{t} = (t_{ij})$ with $i, j$ being s or p. This matrix relates the incident and transmitted electric fields by $\mathbf{E}_{\text{out}} = \underline{t}\mathbf{E}_{\text{inc}}$. In the first step, a THz transmission measurement allows us to determine $t_{\text{pp}}$, relating a p-polarized incoming THz electric field to the p-polarized outgoing THz electric field. In the next step, a THz AHE measurement with a p-polarized incoming THz electric field and a polarizer behind the sample oriented at 45° with respect to the p-direction allows us to infer $t_{\text{sp}} = t_{\text{pp}} \Delta S / S_0$ (see Equation (1) in the main text). Finally, with the help of a $4 \times 4$-transfer-matrix formalism,[11] we establish a relation between $\underline{t}$ and the in-plane conductivity tensor $\underline{\sigma}$, which is solved for $\underline{\sigma}$ numerically. The resulting physically meaningful solution is unique because our samples are optically much thinner than the involved THz wavelengths. In the data extraction procedure, we neglect the possible impact of interface resistances.

***DyCo5.*** In our polycrystalline DyCo5 sample, the DyCo5 crystallites have a hexagonal crystal structure with the c-axis oriented in the sample plane.[50] Therefore, the in-plane conductivity is an average of the ordinary and the extraordinary contribution. We verified the isotropic in-plane conductivity by rotating the DyCo5 sample around the sample normal and found no impact on the measured conductivities.

***CoFe.*** For the CoFe sample, we decided to only determine the average diagonal conductivity of the CoFe(1 nm)|Ta(8 nm)||Si3N4 stack because referencing to a Ta(8 nm)||Si3N4 sample was not possible. The reason is that the sample preparation includes annealing at 300°C, which causes the boron to diffuse from the $Co_{40}Fe_{40}B_{20}$ layer into the Ta layer. As a consequence, the Ta layer changes its chemical composition and optical properties. However, for the off-diagonal conductivity of CoFe, no such complications arose because referencing is unnecessary in this case.

***Ta.*** The conductivity of the Ta layer (total thickness of 8 nm) in the GdFe and DyCo5 samples was determined using a separate Ta(8 nm)||Si3N4(150 nm) stack. The diagonal conductivity extracted from DC and THz measurements is shown in Figure S2, and the corresponding Drude-formula fit parameters are given in Table 1 in the main text. We find a good match between DC and THz results.

***Si3N4.*** The dielectric function of Si3N4 can be described by a superposition of five Lorentzians.[12] By using these literature values, all extracted metal conductivities show unexpected spectral features around 24 THz in the diagonal conductivity. It is known that $Si_3N_4$ has three pronounced phonon resonances in this frequency range.[79] A possible explanation is that during the growth of the metal films on top of the only 150-nm-thick $Si_3N_4$ membrane, a significant amount of strain is induced in the membrane, which can cause changes in its optical constants.

To correct for this effect, the central frequency $\omega_{T_j}$ and the broadening $\Gamma_j$ of the three Lorentzians in this spectral range and the dielectric constant for large frequencies $\varepsilon_\infty$ of the $Si_3N_4$-substrate are slightly adapted (see Table S1). In this way, the most-likely strain-induced spectral feature around 24 THz in the metal conductivities is minimized. We note that for samples fabricated in different laboratories (Berlin and Greifswald), the $Si_3N_4$ optical parameters were adapted separately.

**S5 DC AHE measurements**

The static measurements of $\sigma_{xx}$ and $\sigma_{xy}$ were performed on the same samples used in the THz experiments. For this purpose, the Ta|DyCo$_5$||Si$_3$N$_4$ and Ta||Si$_3$N$_4$ samples were patterned into Hall bar structures by electron-beam lithography and dry-etching steps. A DC current of $I_0 = 200$ µA is passed along the Hall bar structure while the voltages $V_0$ and $\Delta V$ are recorded. A typical measurement of $\Delta V$ is shown in Figure S3.

The longitudinal and transverse resistivities are determined by $\rho_{xy} = \Delta V d/I_0$ and $\rho_{xx} = V_0 b d/I_0 l$ where the width of the Hall bar is $b = 0.97$ µm, its length is $l = 494$ µm and $d$ is the thickness of the magnetic material (Fig. 1a). The resistivities are converted into conductivities via $\sigma_{xx} = \rho_{xx}/(\rho_{xy}^2 + \rho_{xx}^2)$ and $\sigma_{xy} = \rho_{xy}/(\rho_{xy}^2 + \rho_{xx}^2)$. As in the THz experiments, the contribution of the Ta layer is separated using a Ta(8 nm)||Si$_3$N$_4$(500 µm) reference sample. Importantly, because of the requirements for microstructuring, these DC measurements are performed on a part of the sample with a 500-µm-thick Si$_3$N$_4$. Discrepancies in the coercivity field between DC and optical hysteresis curves (compare Figure 2 and Figure S3) may originate from the different substrate thicknesses (150 nm vs 500 µm) used in these two measurements or from different anisotropies caused by microstructuring. This scenario appears reasonable since the magnetic anisotropy and, thus, the coercive field depend sensitively on the strain state of the material.[13]

**S6 Details on the *ab-initio* calculations**

To obtain the frequency-dependent AHE conductivity $\sigma_{yx}(\omega)$, we numerically computed Equation (7) in the main text. The Bloch energies $\epsilon_{kn}$ and Bloch states $|n\mathbf{k}\rangle$ were calculated with a relativistic density-functional theory implementation,[66] in which the spin-orbit interaction is included self-consistently. For the exchange-correlation functional, the local spin-density approximation in the parametrization of von Barth and Hedin was used[14]. The CoFe alloy was modeled as Co$_{0.5}$Fe$_{0.5}$ in the AuCu structure.

## S7 External and proper conductivity

***Definition.*** The Kubo formula determines the conductivity tensor $\underline{\sigma}^{\text{ext}} = (\sigma_{ij}^{\text{ext}})$, which in frequency space relates a perturbing "external" electric field $\mathbf{E}^{\text{ext}}$ to the resulting induced charge-current density $\mathbf{j}$ through[67]

$$\mathbf{j}(\mathbf{x},\omega) = \int d^3\mathbf{x}'\, \underline{\sigma}^{\text{ext}}(\mathbf{x},\mathbf{x}',\omega)\mathbf{E}^{\text{ext}}(\mathbf{x}',\omega) =: (\hat{\sigma}^{\text{ext}}\mathbf{E}^{\text{ext}})(\mathbf{x},\omega). \tag{8}$$

Here, the operator $\hat{\sigma}^{\text{ext}}$ represents the most general linear relationship between $\mathbf{j}$ and $\mathbf{E}^{\text{ext}}$, which also allows for a spatially nonlocal response: A field at position $\mathbf{x}'$ can induce a current density at a different position $\mathbf{x}$. The perturbing field $\mathbf{E}^{\text{ext}}$ can, for instance, be generated by charges on an ungrounded metal plate (plate capacitor) or an electromagnetic emitter. It does, however, not contain the reaction field generated by $\mathbf{j}$. The electric field that shows up in the anomalous velocity of the acceleration theorem of Bloch-state wave packets is also an externally generated electric field.

However, for the constitutive relation found in the Maxwell's equations, a different relationship is used,

$$\mathbf{j}(\mathbf{x},\omega) = \int d^3\mathbf{x}'\, \underline{\sigma}(\mathbf{x},\mathbf{x}',\omega)\mathbf{E}(\mathbf{x}',\omega) =: (\hat{\sigma}\mathbf{E})(\mathbf{x},\omega). \tag{9}$$

It relates the induced current $\mathbf{j}$ with the total electric field $\mathbf{E}$ through the proper conductivity tensor $\underline{\sigma} = (\sigma_{ij})$. Note in Ref. 15 (Section II), it is argued that several software packages (e.g. ELK) do not calculate $\underline{\sigma}^{\text{ext}}$ but rather $\underline{\sigma}$ as these take the mean field into account that is produced by all electrons.

***Relation for thin films.*** The two conductivities can be connected by $\hat{G}_0$, the retarded electromagnetic Green's function operator of free space, which relates any given current density to the resulting electric field. An example is the reaction field $\mathbf{E} - \mathbf{E}^{\text{ext}} = \hat{G}_0\mathbf{j}$ resulting from the induced current $\mathbf{j}$, where

$$(\hat{G}_0\mathbf{j})(\mathbf{x},\omega) = \int d^3\mathbf{x}'\, \underline{G}_0(\mathbf{x},\mathbf{x}',\omega)\mathbf{E}(\mathbf{x}',\omega) \tag{10}$$

with the Green's function $\underline{G}_0(\mathbf{x},\mathbf{x}',\omega)$. By applying the proper conductivity operator $\hat{\sigma}$ to the reaction field above and using Equation (9), we obtain $\mathbf{j} = \hat{\sigma}\mathbf{E}^{\text{ext}} + \hat{\sigma}\hat{G}_0\mathbf{j}$, which along with Equation (8) yields the Dyson-type equation[67]

$$\hat{\sigma}^{\text{ext}} = \hat{\sigma} + \hat{\sigma}\hat{G}_0\hat{\sigma}^{\text{ext}}. \tag{11}$$

Here, it is sufficient to consider an external THz field that is normally incident onto a multilayer sample stacked along the $z$ axis. Therefore, the only spatial dependence is with respect to $z$. For example, the Green's function $\underline{G}_0(z,z',\omega)$ reduces to the scalar $g_0(z,z',\omega)$. Its wave equation is given by[51]

$$(\partial_z^2 + \beta_0^2)g_0(z,z',\omega) = -Z_0 i\beta_0 \delta(z-z') \tag{12}$$

where $\beta_0 = \omega/c$, and $Z_0 \approx 377\,\Omega$ is the vacuum impedance. The retarded solution of Equation (12) is[51] $g_0(z,z',\omega) = -(Z_0/2)\exp(i\beta_0|z-z'|)$.

If the multilayer is much thinner than the wavelength and attenuation length of the THz field inside, both $\mathbf{E}^{\text{ext}}$ and the Green's function can be assumed to be spatially constant throughout the film, with $\mathbf{E}^{\text{ext}}(z,\omega) = \mathbf{E}^{\text{ext}}(\omega)$ and $g_0 = -Z_0/2$. The first term of Equation (11) applied to $\mathbf{E}^{\text{ext}}$ becomes $(\hat{\sigma}^{\text{ext}}\mathbf{E}^{\text{ext}})(z,\omega) = \left(\int dz'\, \underline{\sigma}^{\text{ext}}(z,z',\omega)\right)\mathbf{E}^{\text{ext}}(\omega)$, and similar expressions result for the remaining terms. We $z$-integrate Equation (11) over the multilayer thickness to obtain the matrix equation

$$\underline{H}^{-1} = \left(\underline{H}^{\text{ext}}\right)^{-1} + g_0 \tag{13}$$

in which $\underline{H}(\omega) = \int dz \int dz'\, \underline{\sigma}(z,z',\omega)$ and $\underline{H}^{\text{ext}}(\omega) = \int dz \int dz'\, \underline{\sigma}^{\text{ext}}(z,z',\omega)$ can be considered as conductance tensors of the stack.

We finally assume that $\underline{H} = H_0 + \Delta\underline{H}$ (and, analogously, $\underline{H}_{\text{ext}}$) are dominated by an isotropic (scalar) conductance $H_0$. The small off-diagonal component $\Delta\underline{H}$ can be considered as a perturbation, which may, for example, arise from the AHE. By linearizing $(H_0 + \Delta\underline{H})^{-1} \approx H_0^{-1} - H_0^{-2}\Delta\underline{H}$ in Equation (13), we obtain

$$H_0^{\text{ext}} = \frac{H_0}{1 - H_0 g_0} \tag{14}$$

for the isotropic part, while the off-diagonal part yields

$$\Delta\underline{H}^{\text{ext}} = \frac{\Delta\underline{H}}{(1 - H_0 g_0)^2}. \tag{15}$$

For a homogenous film with thickness $d$ and a single constant conductivity $\underline{\sigma}(\omega)$ and, analogously, $\underline{\sigma}^{\text{ext}}(\omega)$, we have $\underline{H} = \underline{\sigma}d$ and, thus,

$$\Delta\sigma_{xy}^{\text{ext}} = \frac{\Delta\sigma_{xy}}{(1 + Z_0 d\sigma_{xx}/2)^2}. \tag{16}$$

**S8 Simple model of the skew-scattering contribution**

*Diagonal conductivity.* To derive the contribution of skew scattering to the conductivity, we consider the following simple model situation: A pulse $\mathbf{u}_x \delta(t)$ of a homogeneous total electric field acts on the electrons of a solid. At time $t = 0^+$, its action has induced the occupation of Bloch states with band velocity $v_F \mathbf{u}_x$ by $\Delta N_{x0}$ electrons per volume, where $v_F$ denotes the Fermi velocity. Due to subsequent scattering processes, the number of these electrons decays. In the relaxation-time approximation, the decay is given by

$$\Delta N_x(t) = \Delta N_{x0} \Theta(t) \exp(-\Gamma t) \tag{17}$$

with the characteristic time $\Gamma^{-1}$. Because the resulting current density is proportional to $\Delta N_x(t) v_F \mathbf{u}_x$, the conductivity component is in the time domain determined by

$$\sigma_{xx}(t) = \sigma_{\text{DC}} \Gamma \Theta(t) \exp(-\Gamma t), \tag{18}$$

which in the frequency domain yields the Drude formula

$$\sigma_{xx}(\omega) = \frac{\sigma_{DC}}{1 - i\omega/\Gamma}. \tag{19}$$

**Skew-scattering contribution.** The previous treatment implies that in the time interval $[t_0, t_0 + dt_0]$, a fraction $\Gamma dt_0$ of the $\Delta N_x(t_0)$ forward-propagating electrons undergoes scattering. When the sample is magnetized along the $z$ axis and the electric field is applied along the $x$ axis as in Figure 1, electrons have a spin parallel or antiparallel to the $z$ axis. As a consequence, spin-orbit coupling causes a net fraction of the scattered electrons to be deflected along the $y$ direction, thereby acquiring a velocity $v_F \mathbf{u}_y$. In the interval $[t_0, t_0 + dt_0]$, the fraction of these skew-scattered electrons is given by $p_{sk}\Gamma dt_0$, where $|p_{sk}| < 1$ quantifies the strength of skew scattering and, thus, of SOI.

Once the $dN_y(t_0) = \Delta N_x(t_0) \cdot p_{sk}\Gamma dt_0$ electrons have been deflected, they propagate along the $y$ axis, but their number decays with time constant $\Gamma^{-1}$ due to subsequent collisions. Therefore, the number $dN_y(t|t_0)$ of electrons skew-scattered in the interval $[t_0, t_0 + dt_0]$ evolves according to

$$\begin{aligned} dN_y(t|t_0) &= dN_y(t_0) \cdot \Theta(t - t_0) \exp(-\Gamma(t - t_0)) \\ &= \Theta(t_0)\Theta(t - t_0) \exp(-\Gamma t) p_{sk}\Gamma dt_0. \end{aligned} \tag{20}$$

By integrating over all scattering moments $t_0$, we obtain the number $\Delta N_y(t)$ of the currently propagating skew-scattered electrons. The result

$$\Delta N_y(t) = p_{sk}\Gamma t \cdot \Theta(t) \exp(-\Gamma t) = p_{sk}\Gamma t \cdot \Delta N_x(t). \tag{21}$$

can be interpreted as follows: At time $t = 0$, the electric-field pulse generates a group of forward-propagating electrons. As time passes, the number of skew-scattered electrons increases linearly ($\propto p_{sk}\Gamma t$). Simultaneously, however, the total number $\Delta N_x$ of still forward-propagating electrons decreases exponentially ($\propto \exp(-\Gamma t)$).

From Equation (21), we obtain the skew-scattering contribution to the AHE conductivity in the time domain, which simply reads

$$\sigma_{yx}^{sk}(t) = p_{sk} t \sigma_{xx}(t). \tag{22}$$

A Fourier transformation yields $\sigma_{yx}^{sk}$ in the frequency domain,

$$\sigma_{yx}^{sk}(\omega) = p_{sk}\frac{\sigma_{DC}}{(1 - i\omega/\Gamma)^2}. \tag{23}$$

We note that the frequency dependence of $\sigma_{yx}^{sk}(\omega)$ is identical to that of the ordinary Hall effect.[16] For frequencies $\omega \ll \Gamma$, $\sigma_{yx}^{sk}(\omega)$ grows linearly with the strength of SOI ($\propto p_{sk}$) and the DC conductivity ($\sigma_{DC} \propto \Gamma^{-1}$). Therefore, skew scattering is only relevant in samples with large $\sigma_{DC}$. We emphasize that the scaling $\sigma_{yx}^{sk}(\omega = 0) \propto \Gamma^{-1}$ is in full agreement with the formal definition of skew scattering.[5]

**Table S1 | Optical constants used for Si$_3$N$_4$.** Constants used for the refractive index calculation Si$_3$N$_4$ (taken from Ref. 79). Slight adaptions are necessary due to possibly strain-induced modifications during sample growth (see text for details). For the samples prepared at the Helmholtz-Zentrum Berlin (Ta, DyCo$_5$ and GdFe) different constants are used than for the sample prepared at the University of Greifswald (CoFe(1 nm)|Ta(8 nm)).

| Parameter | Literature value [Cataldo, et al.] | Constants for Ta, DyCo$_5$ and GdFe | Constants for CoFe(1)|Ta(8) |
|---|---|---|---|
| $\omega_{T3}/2\pi$ (THz) | 24.52 | 23.98 | 23.82 |
| $\omega_{T4}/2\pi$ (THz) | 26.44 | 26.06 | 26.11 |
| $\omega_{T5}/2\pi$ (THz) | 31.72 | 29.73 | 30.38 |
| $\Gamma_3/2\pi$ (THz) | 2.75 | 4.43 | 2.96 |
| $\Gamma_4/2\pi$ (THz) | 3.48 | 4.11 | 3.55 |
| $\Gamma_5/2\pi$ (THz) | 5.95 | 2.97 | 5.53 |
| $\varepsilon_\infty$ | 4.56+0.01i | 4.36+2.53i | 5.62+1.43i |

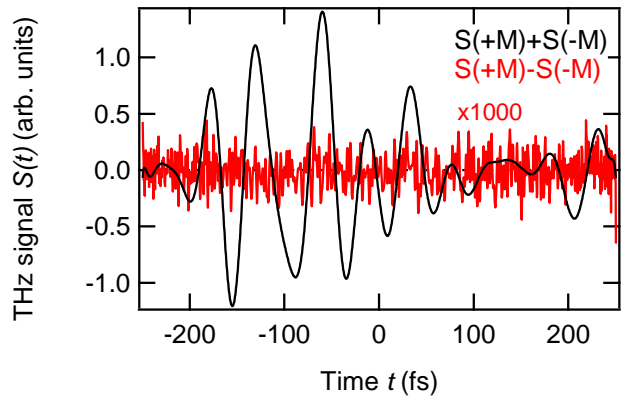

**Figure S1.** Reference measurement using a bare Si$_3$N$_4$ membrane without metallic layers.

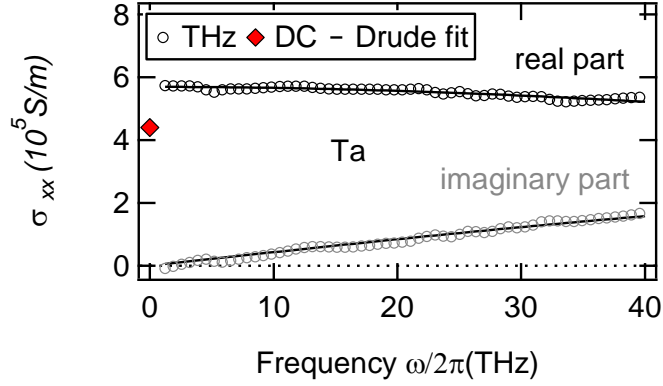

**Figure S2.** Diagonal conductivity σ$_{xx}$ of tantalum measured in the THz frequency range (real part: black crosses, imaginary part: red crosses) and at DC (diamond symbol). The fit (solid lines) was obtained using the Drude model (see Equation (6) in the main text and Table 1).

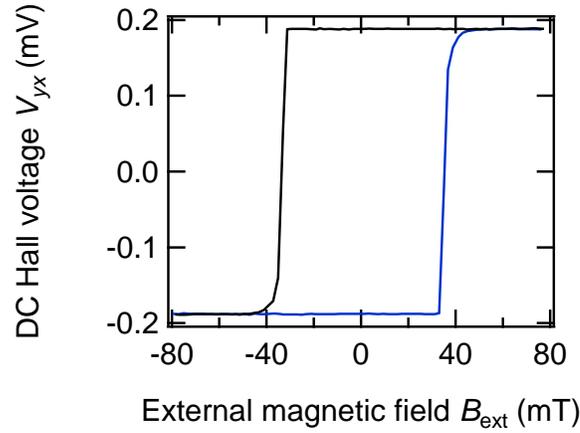

**Figure S3. Electrical anomalous Hall measurement of DyCo$_5$.** DC anomalous Hall voltage $V_{yx}$ vs the out-of-plane-oriented external magnetic field **B**$_{ext}$ measured electrically on a microstructured DyCo$_5$ sample.

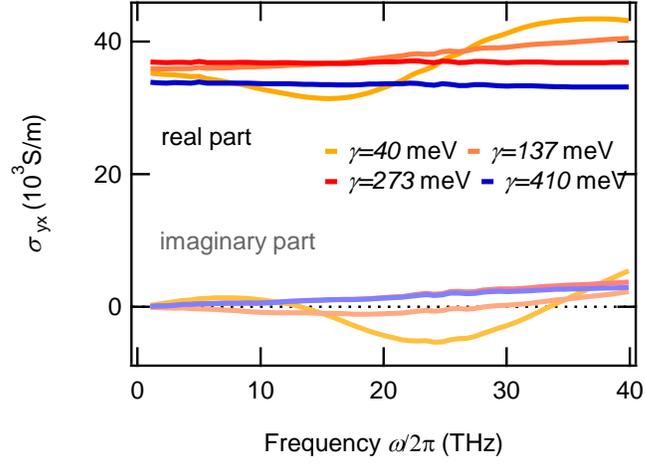

**Figure S4. *Ab-initio*-calculated AHC for different broadenings.** Theoretical off-diagonal conductivity $\sigma_{yx}^{calc}$ of CoFe calculated for different broadenings according to Equation (7) in the main text, taking into account the proper conductivity (see Supporting Information S7). The light-colored lines are the respective imaginary parts.